\documentclass[11pt,twoside]{article}

\usepackage{asp2006}
\usepackage{lscape}
\usepackage{graphicx}

\begin{document}
\title{Photometric search of orbital periods in symbiotic stars}

\author{M.~Wi\c{e}cek$^{1}$, M.~Miko\l{}ajewski$^{1}$, T.~Tomov$^{1}$, M.~Cika\l{}a$^{1}$, C.~Ga\l{}an$^{1}$, A.~Majcher$^{1,2}$, E.~\'Swierczy\'nski$^{1}$, P.~Wychudzki$^{1}$, P.~R\'o\.zanski$^{1}$, S.~Fr\c{a}ckowiak$^{1}$, J.~L.~Janowski$^{1}$, D.~Graczyk$^{1}$}

\affil{$^{1}$Centre for Astronomy of Nicolaus Copernicus University, Toru\'n, Poland}
\affil{$^{2}$Soltan Institute for Nuclear Studies, Warsaw, Poland}

\begin{abstract} 
We present UBVRI photometry of three symbiotic stars ZZ CMi, TX CVn and AG Peg carried out  from 1997 to 2008 in Piwnice Observatory near Toru\'n. To search orbital periods of these stars Fourier analysis was used. For two of them, TX CVn and AG Peg, we have confirmed the earlier known periods. For ZZ CMi we found a relatively short period 218.59~days. Assuming, that the orbital period is twice longer (P=437.18~days), the double sine wave in the light curve can be interpreted by ellipsoidal effect.
\end{abstract}

\section*{Photometric data}
We used a 60cm Cassegrain telescope equipped with  an EMI~9558B photomultiplier (1991~--~1999), a RCA~C31034 photomultiplier (2001~--~2004) and CCD camera (2005~-~2008), to obtain the $UBVRI$ photometry of the selectd symbiotic stars \citep[for details see][]{galan}. For TX CVn and AG Peg we used additional data from the Slovak long-term monitoring programme of photometric observations of selected symbiotic stars \citep[and references therein]{hric,skopal}.

\section*{ZZ Canis Minoris}

\begin{figure}[!htp]
  \centering  
  \includegraphics[width=0.5\textwidth, angle=270]{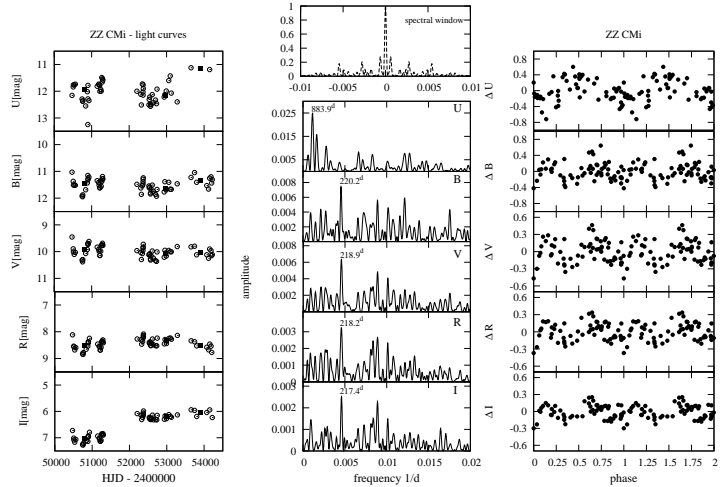}
  \caption{Left panel: Multicolour light curves of ZZ CMi. Black squares are
mean values of brightness to normalize three photometric systems for Fourier analysis,
Middle panel: Power spectrum for $UBVRI$ filters data of ZZ~CMi,
Right panel: The residual $BVRI$ and U light curves of ZZ~CMi phased
with 437.18 and  835.12 days respectively. Residual magnitudes were obtained by
substraction of a mean value from observational data (see Figure~\ref{zzcmi}, left panel) for each of 
the photometric systems.}
  \label{zzcmi}
\end{figure}

All multicolour photometric measurements of  ZZ~CMi are presented in Figure~\ref{zzcmi} (left panel). The Fast Fourier Transform (FFT) of the data gives a 218.59~days period for $BVRI$ filters and 835.12 days~--~for U (Figure~\ref{zzcmi} middle panel). On the right panel of Figure~\ref{zzcmi} the $BVRI$ light curves phased with the double period (437.18~days) are demonstrated.

We have used Wilson~--~Devinney's code \citep{wilson} to verify the assumption, that the double sine wave could be interpret as an ellipsoidal effect connected with the tidal distortion of the giant surface. A circular orbit ($e=0$) with inclination $i=90^{o}$ was adopted. We have determined the ephemeris $JD_{\phi_{0}}=2450326.24 + 437.18 \times E$, and used it to phase the data. The temperature of the cool component was assumed to be $T=3400~K$ according to M5III classification of \citet{taranova}. Many solutions for the components mass, in the range $1-3~M_{\odot}$ for the giant and $0.5-1.4~M_{\odot}$ for the white dwarf, were tested.

\begin{figure}[!b]
  \begin{center}  
    \includegraphics[scale=0.235, angle=0]{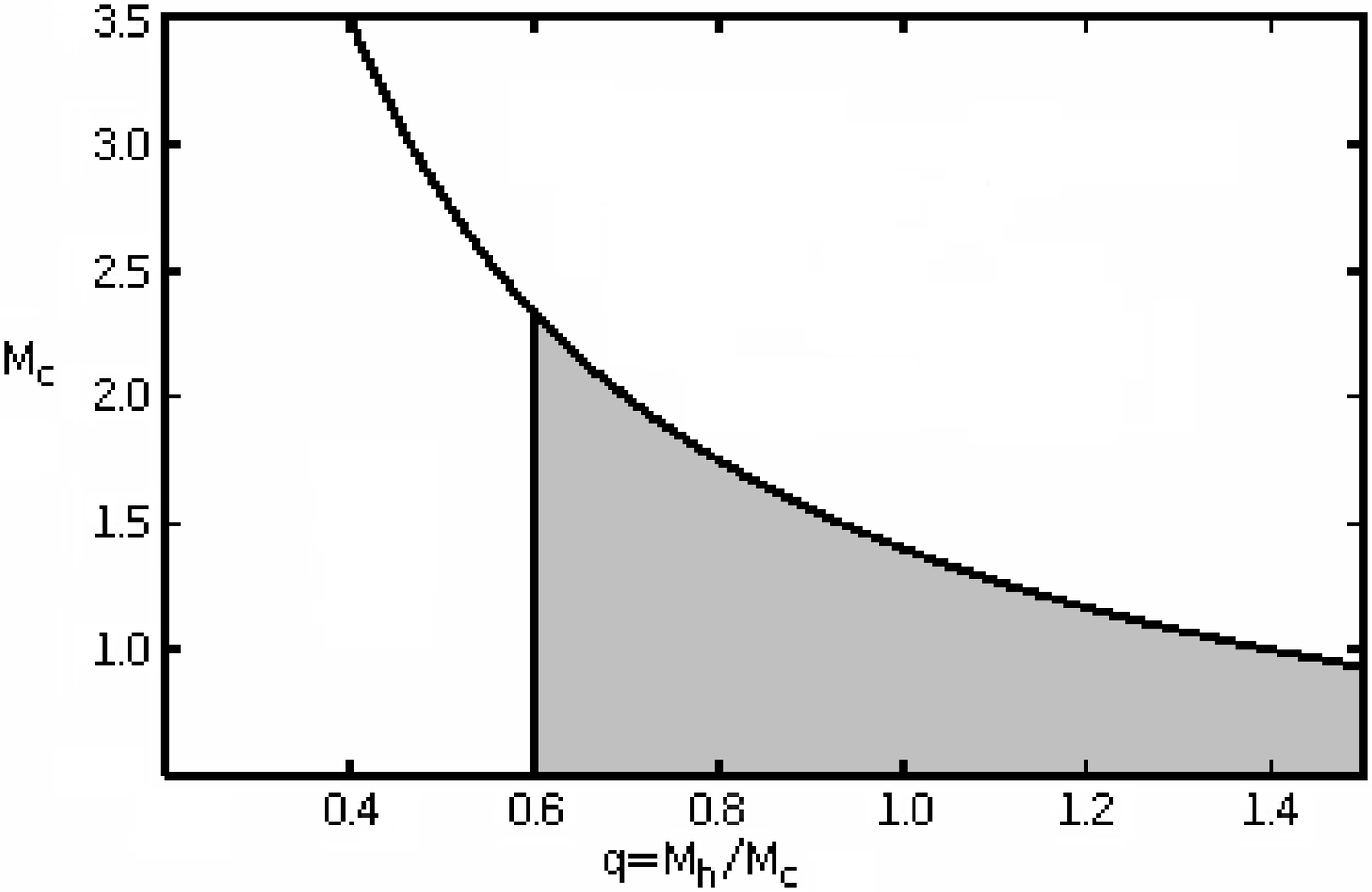}\hspace*{1cm}
    \includegraphics[scale=0.42, angle=0]{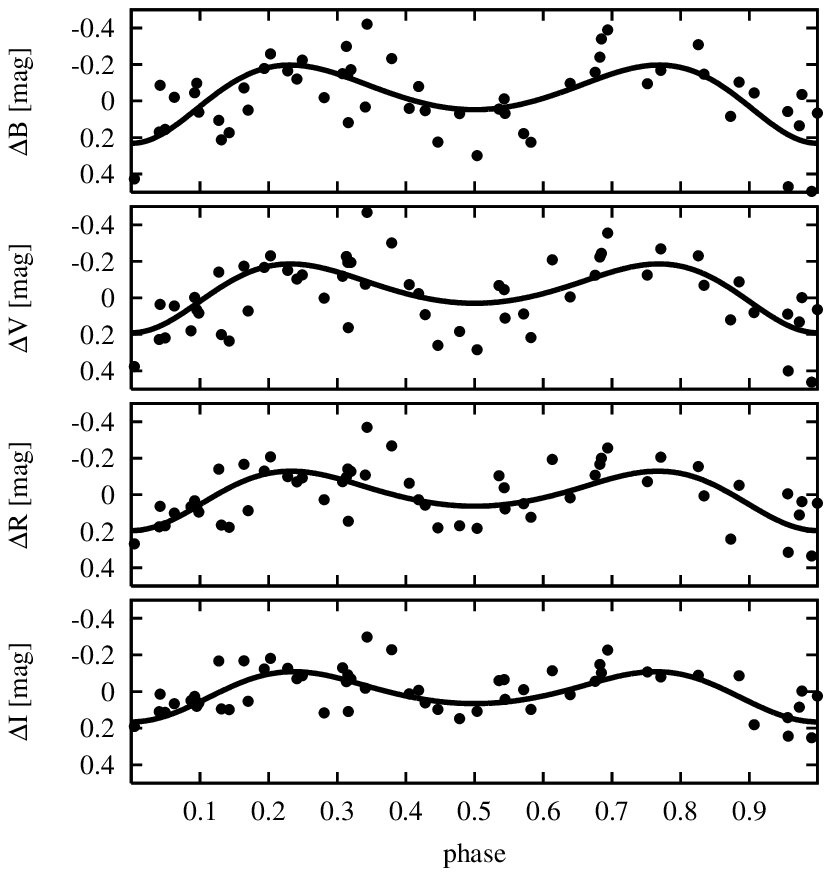}
  \end{center}
  \caption{Left panel: the relationship between $q = M_{h}/M_{c}$
  and $M_{c}$ is presented. The filled area shows the possible mass solutions for which
  the cool giant does not overfill the Roche surface and the mass of the
  white dwarf is smaller than $1.4M_{\odot}$, Right panel: The $BVRIC$ light curves phased with double of
  218.59 days period. Solid lines represent the synthetic curves computed for  
  $M_{c} = 1.0M_{\odot}$, $M_{h} = 1.0M_{\odot}$.}
\label{zzcmims}
\end{figure}

The possible solutions, in which the hot star is less massive than $1.4~M_{\odot}$ and the giant doesn't overflow Roche lobe, are restricted to the filled area in Figure~\ref{zzcmims} (left panel). It is clearly seen that the cool component has to be less massive than $\sim 2M_{\odot}$ and the mass ratio (q) is greater than $\sim 0.6$. All synthetic light curves obtained for combination of masses from the filled region in Figure~\ref{zzcmims} (left panel) give practically the same solutions, similar to that shown in Figure~\ref{zzcmims} (right panel).

\section*{TX Canum Venaticorum}

TX CVn is symbiotic star, which spectrum is combined by a late B and an early M star spectra \citep{kenyon86} and with an orbital  period  $199 \pm 3$ days \citep{garcia}. Spectrophotometric observations shows that the hot component is not a normal late {\it B-type}  star, because the He~I features are produced in an expanding with a velocity of $\sim 300 km\,s^{-1}$ photosphere.

\begin{figure}[!b]
\begin{center}  
\includegraphics[scale=0.38, angle=270]{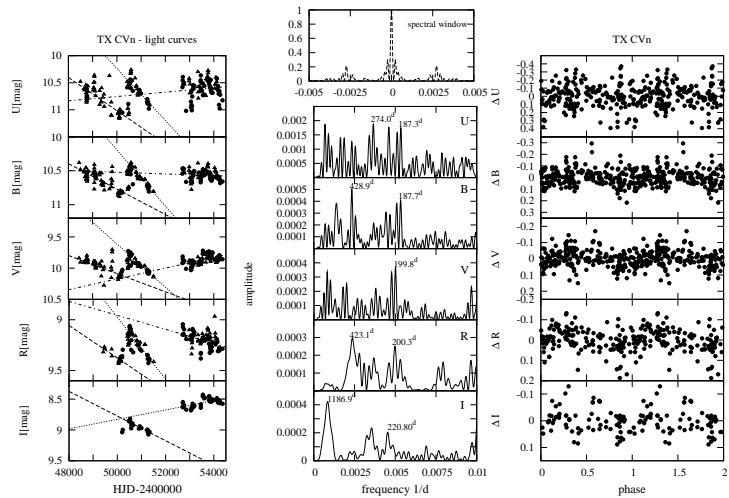}
\end{center}
\caption{Left panel: Multicolour light curves of TX~CVn with the lines fitted
to subtract the trends.
Middle panel:  The power spectrum for $UBVRI$ filters data of TX~CVn.
Right panel: The $UBVRI$ light curves of TX~CVn -- the residual data after the 
trend subtraction phased with the 199.75 days period. The $I_c$ light curve shows evidence for an elliptical effects.
The triangles mark the Slovak data, the circles - our data.}
\label{txcvn}
\end{figure}

We used photometric data to confirm the spectroscopic period of this system. To subtract the long term trends we fitted lines to
our data (Figure~\ref{txcvn} left panel), and then the FFT was calculated separately for the {\it UBVRI} data. The power spectra are shown in the middle panel of Figure~\ref{txcvn}. The periodograms show a peak at frequency $f\sim0.0050\,d^{\mathrm{-1}}$, which corresponds to the known period of $\sim 199$ days. The phased light curves are shown in Figure~\ref{txcvn} (right panel).

\section*{AG Pegasi}

AG~Peg is a symbiotic binary consisting of a white dwarf and a red giant. The stars orbiting their mass center with a period of $812.3 \pm 6.3$ days \citep{kenyon93}. \citet{murset} found the evidence for colliding stellar winds in this system: a fast wind ($V \sim 900\,km\,s^{-1}$) from the hot component and a slow wind ($V \sim 60\,km\,s^{-1}$) from the giant. The colliding winds form the gaseous nebula surrounding the system.

\begin{figure}[htp]
  \begin{center}  
	\includegraphics[scale=0.37, angle=270]{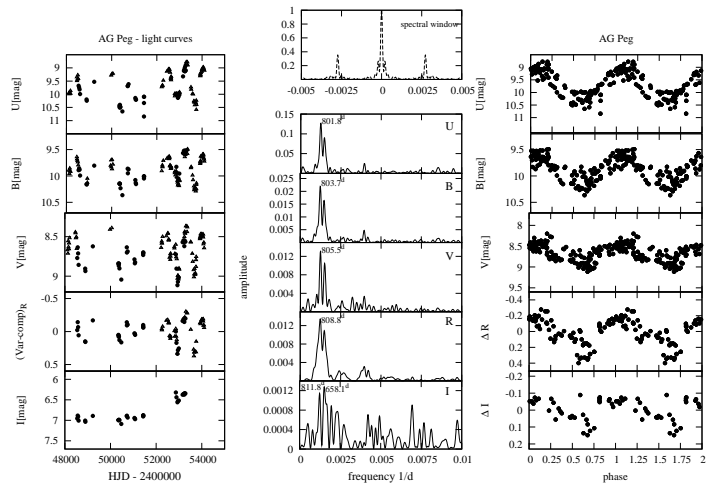}
  \end{center}
  \caption{ Left panel: Multicolour light curves of AG Peg. The open triangles mark the
  Slovak data, the circles - our data. Middle panel: Power spectrum for
  $UBVRI$ filters data of AG~Peg, Right panel: The $UBVRI$ light curves of AG Peg phased
  with 806.31 days period. In the two lower panels $\Delta R=R_{var} - R_{comp}$ and $\Delta I=I_{var} - I_{comp}$ are used.}
\label{agpeg}
\end{figure}

Our $UBVRI$ photometric measurements of this star are presented in Figure~\ref{agpeg} (left panel). The periodograms of multicolour photometry are shown in  Figure~\ref{agpeg} (middle panel). We adopted a mean value of 806.31~days for the period in all filters obtained from the FFT analysis. This value is within the error box of the period published by  Kenyon et al. (1993). The data were phased with our period, using $T_{0} = 2442710.1$ from \citet{fernie}. The phased light curves are shown in Figure~\ref{agpeg} (right panel).

\acknowledgements This work is supported by the Polish MNiSW Grant N203~018~32/2338.

\end{document}